\documentclass[submission,copyright,creativecommons]{eptcs}
\providecommand{\event}{SYNT'14} 

\usepackage{latexsym,amsmath,amssymb}
\usepackage{graphics,graphicx,color}
\usepackage{gnuplot-lua-tikz}
\usepackage{comment}
\usepackage{tikz}

\def\R{\mathbb R}

\def\N{\mathbb N}

\def\X{\mathbb X}

\def\la{\lambda}        
\def\si{\sigma}

\def\cC{\mathcal C}             \def\cD{\mathcal D}
\def\cE{\mathcal E}             \def\cF{\mathcal F}
             \def\cH{\mathcal H}
             
             \def\cL{\mathcal L}

             \def\cV{\mathcal V}
\def\cW{\mathcal W}

\def\bA{\mathbf A}              \def\bB{\mathbf B}
              
\def\bE{\mathbf E}              \def\bF{\mathbf F}
\def\bG{\mathbf G}              
\def\bI{\mathbf I}

              \def\bP{\mathbf P}
              
\def\bS{\mathbf S}              \def\bT{\mathbf T}
\def\bU{\mathbf U}

\def\ip#1{\left<#1\right>}

\def\diag{\mbox{diag}}

\newcommand{\syndef}{\mbox{\tt::=}}
\newcommand{\synalt}{\;\mbox{\tt\large$|$}\;}
\newcommand{\sem}[1]{[\![ #1 ]\!]}

\newcommand{\tup}[1]{\langle #1 \rangle}

\newcommand{\pp}[1]{\mbox{${}^{#1}$}}

\newcommand{\init}{\mbox{\sl init}}

\newcommand{\States}{\mbox{\bf State}}
\newcommand{\ProbStates}{\mbox{\bf ProbState}}
\newcommand{\Confs}{\mbox{\bf Conf}}

\newcommand{\Dists}{\mbox{\bf Dist}}

\newcommand{\Values}{\mbox{\bf Value}}

\newcommand{\Vars}{\mbox{\bf Var}}

\newcommand{\Labs}{\mbox{\bf Lab}}

\newcommand{\false}{\mbox{\tt false}}
\newcommand{\true}{\mbox{\tt true}}

\newcommand{\varE}[1]{\mbox{\tt #1}}

\newcommand{\skipS}{\mbox{\tt skip}}
\newcommand{\skipL}[1]{[\mbox{\tt skip}]\pp{#1}}

\newcommand{\stopS}{\mbox{\tt stop}}

\newcommand{\getsS}[2]{#1 ~\mbox{\tt :=}~ #2}
\newcommand{\getsL}[3]{[#1 ~\mbox{\tt :=}~ #2]\pp{#3}}

\newcommand{\ranS}[2]{#1 ~\mbox{\tt ?=}~ #2}
\newcommand{\ranL}[3]{[#1 ~\mbox{\tt ?=}~ #2]\pp{#3}}

\newcommand{\chooseS}[2]{\mbox{\tt choose}~#1~\mbox{\tt or}~#2~\mbox{\tt ro}}
\newcommand{\chooseL}[3]{[\mbox{\tt choose}]\pp{#3}~#1~\mbox{\tt or}~#2~\mbox{\tt ro}}

\newcommand{\whileL}[3]{\mbox{\tt while}~[#1]\pp{#3}~\mbox{\tt do}~#2~\mbox{\tt od}}

\newcommand{\ifS}[3]{\mbox{\tt if}~#1~\mbox{\tt then}~#2~\mbox{\tt else}~#3~\mbox{\tt fi}}
\newcommand{\ifL}[4]{\mbox{\tt if}~[#1]\pp{#4}~\mbox{\tt then}~#2~\mbox{\tt else}~#3~\mbox{\tt fi}}

\newcommand{\lsem}[1]{\sem{#1}}
\newcommand{\ksem}[1]{\sem{#1}}

\newtheorem{proposition}{Proposition}
\newtheorem{definition}{Definition}

\title{Program Synthesis and Linear Operator Semantics}
\author{Herbert Wiklicky
\institute{%
Department of Computing,
Imperial College London,
London, United Kingdom}
\email{herbert@imperial.ac.uk}
}

\def\titlerunning{Program Synthesis and Linear Operator Semantics}
\def\authorrunning{H. Wiklicky}

\begin{document}
\maketitle

 \begin{abstract}
   For deterministic and probabilistic programs we investigate the problem of
   program synthesis and program optimisation (with respect to non-functional
   properties) in the general setting of global optimisation. This approach is
   based on the representation of the semantics of programs and program
   fragments in terms of linear operators, i.e. as matrices. We exploit in
   particular the fact that we can automatically generate the representation of
   the semantics of elementary blocks. These can then can be used in order to
   compositionally assemble the semantics of a whole program, i.e. the
   generator of the corresponding Discrete Time Markov Chain (DTMC). We also
   utilise a generalised version of Abstract Interpretation suitable for this
   linear algebraic or functional analytical framework in order to formulate
   semantical constraints (invariants) and optimisation objectives (for example
   performance requirements).
 \end{abstract}


\section{Introduction}
\label{Introduction}

The automatic generation or synthesis and optimisation of code is an extremely
complex task, nevertheless it constitutes to some extend the holy grail of
software engineering. In this paper we consider an approach to this problem
via a non-standard semantical model of programs in terms of linear operator
or, simply, as matrices. This allows us to employ well-developed techniques of
classical mathematical (non-linear) optimisation.
More concretely we describe here an experimental implementation of the
framework which treats programs more as dynamical systems than as logical
entities. In this setting we then aim in generating or transforming programs
on the basis of optimising some of their properties. In this way we try to end
up with code that exhibits the desired properties (at least as much as
possible).

The initial motivation of this approach can be traced back to when we
considered Kocher's attack on the RSA algorithm \cite{ICICS08,IJIS11}. In very
simple terms \cite{Kocher96}: the problems is that the execution time of a
certain algorithm (e.g. modular exponentiation) is based on some secrete or
high information (concretely, the bits in a secrete key) and thus it is
possible to reveal or extract the secret by analysing the running time.  For
example we have repeatedly, for each bit $k[i]$ to execute code which takes
very little or a lot of time: $\ifS{k[i]}{\tup{short}}{\tup{long}}$.

In, for example, \cite{Agat00}\ it has been suggested to obfuscate the time
signature by using depleted versions $[short]$ and $[long]$ of $\tup{short}$
and $\tup{long}$, respectively; i.e. code which is executed in the same time
as the original version but which does otherwise not change the state in any
way.  This {\em padding} means that we are replacing
\(
\ifS{k[i]}{\tup{short}}{\tup{long}}
~\mbox{by}~
\ifS{k[i]}{\tup{short};~[long]}{[short];~\tup{long}}.
\) 
The result is then that both branches always take the same maximal time 
to execute. 

As there is a tradeoff between how easy it is to obtain the secrete (key) from
the time signature and the increased running time we suggested to introduce
the extra time randomly. The result is a whole manifold of programs $P(p)$ in
which the padding is performed with a certain probability $p$ or the original
code is executed with probability $1-p$. The idea is now to find the $p^*$ for
which we have the optimal balance between extra cost and security.

The purpose of this paper is to extend this idea to allow the generation or
transformation of programs as an optimisation problem. We will consider a
whole family of programs parameterised by a large number of variables $\la_i$
and try to identify those which fulfil certain requirements in an optimal way.


\section{The General Approach}

Program synthesis goes back to the work by Manna and Waldinger in the late
1960s and 70s. It received renewed interest in the last years, in particular
in the area of protocol and controller synthesis, see e.g. the recent special
issue on ``Synthesis'' \cite{BodikJobstmann13}\ where various approaches
towards program synthesis presented. To some degree our approach is related to
``Program Sketching'' \cite{SolarLezama13}, we only provide a `sketch' of a
program which leaves certain parts (blocks, statements) open. In order to
fulfil a given specification or to meet certain performance objectives one can
employ various algorithms in order to determine the appropriate concrete
statements, chosen from a set of potential, possible implementations. In this
setting one can distinguish between an {\em implementation} language and a
{\em specification} language which allows the description of certain templates
(including valid alternative implementations) and of assertions,
i.e. constraints the implementation should ultimately fulfil.

The central idea of our approach is to interpret the probabilistic choice in a
program not as a choice made at {\em run-time} but as a parameter which
describes a whole manifold of possible programs or perhaps better their
semantics. The aim is the identification of the ``right'' parameters at {\em
  compile-time} and thus to generate or synthesise the desired program
behaviour.

\subsection{A Manifold of Sketches}

Our approach can be also seen as a form of {\em continuous} sketching in the
sense of Solar-Lezama \cite{SolarLezama13}. Instead of allowing for a sketch
like (cf 
\cite[p26]{SolarLezama08})
\begin{verbatim}
int W = 32;
void main(bit[W] x, bit[W] y){
  bit[W] xold = x;
  bit[W] yold = y;
  if(??) { x = x ^ y;} else { y = x ^ y}; }
  if(??) { x = x ^ y;} else { y = x ^ y}; }
  if(??) { x = x ^ y;} else { y = x ^ y}; }
  assert y == xold && x == yold;
}
\end{verbatim} 
our approach we would consider something like
\begin{verbatim}
int W = 32;
void main(bit[W] x, bit[W] y){
  bit[W] xold = x;
  bit[W] yold = y;
  choose p:{ x = x ^ y;} or 1-p:{ y = x ^ y}; } ro
  choose q:{ x = x ^ y;} or 1-q:{ y = x ^ y}; } ro
  choose r:{ x = x ^ y;} or 1-r:{ y = x ^ y}; } ro
  assert y == xold && x == yold;
}
\end{verbatim} 
where the {\tt choose p:S1 or 1-p:S2 ro} statements implements random choices:
With probability $p$ we execute {\tt S1} and with probability $1-p$ we execute
{\tt S2}. The aim is to identify in the design space the correct or optimal
parameters $p,q,r,\ldots\in[0,1]$ such that certain conditions or assertions
are fulfilled.

We will specify the semantics of programs as Linear Operators on an
appropriate vector space (containing all probabilistic states) or simply as
matrices, i.e. $\sem{P}\in\cL(\cV)$, which describe the generator of a
Discrete Time Markov Chain (DTMC). The desired properties -- i.e. the
objectives of a certain synthesis problem -- will be recast as properties of
this linear operator. In this way, the assertions to be fulfilled are
translated into a (non-linear) optimisation problem with the appropriate
objective function $\Phi:\cL(\cV) \rightarrow \R$ and constraints
(e.g. guaranteeing the normalisation of probabilities). The synthesis problem
becomes in this way a global (in general non-linear) optimisation
(minimisation or maximisation) problem.


\subsection{Probabilistic Abstract Interpretation}

In order to relate usual notions of program properties and the objective
function $\Phi$ we will utilise our theory of Probabilistic Abstract
Interpretation (PAI) which generalises Cousot's Abstract Interpretation
framework (though it is different from the approach in, for example,
\cite{Monniaux00,CousotMonerau12}).

Classically the correctness of a program analysis is asserted with respect to
the semantics in terms of a correctness relation. The theory of Abstract
Interpretation allows for constructing analyses that are automatically correct
without having to prove it a posteriori
\cite{CousotCousot77a,CousotCousot79a}.
The main applications of this theory are for the analysis of safety-critical
systems as it guarantees correct answers at the cost of precision.

For probabilistic systems or the probabilistic analysis of (non-)deterministic
ones, the theory of Probabilistic Abstract Interpretation (PAI) allows for the
construction of analyses that are possibly unsafe but maximally precise
\cite{PPDP00,LOPSTR00}.  Its main applications are therefore in fields like
speculative optimisation and the analysis of trade-offs.  PAI has been used
for the definition of various analyses based on the LOS (see
e.g. \cite{QAPL08,APLAS07,TCS04}. In the following we will, for the sake of a
simpler mathematical treatment, only consider finite dimensional versions of
PAI (although it is also possible to extend the framework to infinite
dimensional spaces, e.g. \cite{APLAS13, FOPARA14}).

PAI relies on the notion of generalised (or pseudo-)inverse in place of the
notion of a Galois connection as in classical Abstract Interpretation. This
notion is well-established in mathematics where it is used for finding
approximate, so-called least-square solutions (cf. e.g.
\cite{BenIsraelGreville03}).

\begin{definition}
  Let $\cH_1$ and $\cH_2$ be two Hilbert spaces and $\bA: \cH_1 \mapsto \cH_2$
  a  linear map between them. A linear map $\bA^\dagger =
  \bG: \cH_2\mapsto \cH_1$ is the {\em Moore-Penrose pseudo-inverse} of $\bA$
  iff
  $\bA \circ \bG = \bP_\bA$
  and
  $\bG \circ \bA = \bP_\bG$,
  where $\bP_\bA$ and $\bP_\bG$ denote orthogonal projections onto the ranges
  of $\bA$ and $\bG$.
\end{definition}
An linear operator $\bP:\cH\rightarrow\cH$ is an {\em orthogonal projection}
if $\bP^* = \bP^2 = \bP$, where ${.}^*$ denotes the {\em adjoint}. The adjoint
is defined implicitly via the condition: $\ip{x\cdot\bP,y}=\ip{x,y\cdot\bP^*}$
for all $x,y\in\cH$, where $\ip{.,.}$ denotes the inner product on $\cH$. For
real matrices the adjoint correspond simply to the transpose matrix
$\bP^*=\bP^t$ \cite[Ch~10]{Roman05}.

If $\cC$ an $\cD$ are two Hilbert spaces, and $\bA: \cC \rightarrow \cD$ and
$\bG: \cD \rightarrow \cC$ are linear operators between the concrete
domain $\cC$ and the abstract domain $\cD$, such that $\bG$ is the
Moore-Penrose pseudo-inverse of $\bA$, then we say that $(\cC,\bA,\cD,\bG)$
forms a {\em probabilistic abstract interpretation}.

A very simple example of such a abstraction is the {\em forgetful abstraction}
$\bA_f:\R^n \rightarrow \R$ which is represented by an $n\times 1$ dimensional
(column) matrix with all entries equal to $1$. Its Moore-Penrose
pseudo-inverse $\bA_f^\dagger$ is simply a $1\times n$ (row) matrix with all
entries $\frac{1}{n}$. This abstraction ``forgets'' about all details and just
records the existence of a system (part).


\section{The Language}
\label{Language}

Our approach uses the same language to specify implementations and templates
or sketches. We use a language which allows for a probabilistic (rather than a
non-deterministic) choice. If we utilise this to describe an implementation
the idea is that the choice is made at run time according to a given
probability (by a `coin-flipping' device) while as a specification language
the probabilities are chosen a priori, at compile-time such as to optimise the
behaviour or performance. This could be summarised as: Probabilities are
variables in the context of synthesis and constants when executed. The
objectives of a synthesis tasks can be expressed by any appropriate function
on the space of possible semantics (which in our case has the structure of a
vector space or linear algebra), which we see as a kind of semantical
abstraction.


\subsection{Syntax}

We consider a labelled version of the standard (probabilistic) procedural
language as one can find it for example in \cite{Kozen81}. Further details can
be found, e.g., \cite{Bertinoro10}. All statements are labelled in order to
allow a convenient construction of its semantics (indicating relevant program
points). These labels can always be reconstructed if a unlabelled version of
a program is considered.

\begin{table}[h]
\begin{center}
 \begin{minipage}{5cm}
 \begin{center}
 \begin{tabular}{rcl}
 $S$ & $\syndef$ & $\skipL{\ell}$ \\
     & $\synalt$ & $\getsL{x}{f(x_1,\ldots,x_n)}{\ell}$\\
     & $\synalt$ & $\ranL{x}{\rho}{\ell}$ \\
     & $\synalt$ & $S_1$\verb|;| $S_2$ \\ 
     & $\synalt$ & $\chooseL{p_1:S_1}{p_2:S_2}{\ell}$ \\
     & $\synalt$ & $\ifL{b}{S_1}{S_2}{\ell}$ \\
     & $\synalt$ & $\whileL{b}{S}{\ell}$ \\
 \end{tabular}
 \end{center}
 \end{minipage}
 \end{center}
 \caption{The Labelled Syntax}
 \label{Syntax}
 \end{table}

The statement $\skipS$ does not have any operational effect but can be used,
for example, as a placeholder in conditional statements. We have the usual
(deterministic) assignment  $\getsS{x}{e}$, sometimes also in
the form $\getsS{x}{f(x_1,\ldots,x_n)}$.
In the random assignment $\ranS{x}{\rho}$, the value of a variable $x$ is set
to a value according to some random distribution $\rho$. In \cite{Kozen81}\ it
is left open how to define or specify distributions $\rho$ in detail. We will
use occasionally an ad-hoc notation as sets of tuples $\{(v_i,p_i)\}$
expressing the fact that value $v_i$ will be selected with probability $p_i$;
or just as a set $\{v_i\}$ assuming a uniform distribution on the values
$v_i$. It might be useful to assume that the random number generator or
scheduler which implements this construct can only implement choices over
finite ranges, but in principle we can also use distributions with infinite
support. The statement $\chooseS{p_1:S_1}{p_2:S_2}$ executes randomly
either $S_1$ or $S_2$, assuming an implicit normalisation of probabilities,
i.e. $p_1+p_2=1$. For the rest we have the usual sequential composition,
conditional statement and loop.  We leave the detailed syntax of functions $f$
or expressions $e$ open as well as for boolean expressions or test $b$ in
conditionals and loop statements. For each (labelled) statement in this
language we identify the initial and final label


\subsection{Linear Operator Semantics}

The Linear Operator Semantics (LOS) is intended to model probabilistic
computations we therefore have to consider probabilistic states. These
describe the situation about the computation at any given moment in time. Our
model is based on a discrete time model. The information will specify the
probability that the computational system in question is in a particular
classical state.

A {\em classical state} $s\in\States=\Vars\rightarrow\Values$ associates a
certain value $s(x)$ with a variable $x$. We assume, in order to keep the
mathematical treatment as simple as possible, that the possible values are
finite, e.g. $\Values=\{-MININT,\ldots,MAXINT\}$.
A {\em probabilistic state} $\si\in\ProbStates=\States\rightarrow[0,1]$
can be seen as a probability distribution on classical states or as a 
(normalised) vector in the free vector space $\cV(\Values)$ over $\Values$.

The set of probabilistic states forms a (sub-set) of a (finite-dimensional)
vector (Hilbert) space. The semantics $\bT(P)=\sem{P}$ of a program $P$ is a
linear map or operator on this vector space which encodes the generator of a
Discrete Time Markov Chain (DTMC). DTMC are non-terminating processes: it is
assumed that there is always a next state and the process goes on forever. In
order to reflect this property in our semantics, we introduce a terminal
statement {\tt stop} which indicates successful termination. Then the
termination with a state $s$ in the classical setting is represented here by
reaching the final configuration $\tup{{\tt stop},s}$ which then `loops'
forever after. This means that we implicitly extend a statement $S$ to
construct full programs of the form $P \equiv S;~[{\tt stop}]^{\ell^*}$.

The {\em tensor product} is an essential element of the description of
probabilistic states and the semantical operator $\bT(P)$. The tensor product
-- more precisely, the Kronecker product, i.e.  the coordinate based version
of the abstract concept of a tensor product -- of two vectors
$(x_1,\ldots,x_n)$ and $(y_1,\ldots,y_m)$ is given by
$(x_1y_1,\ldots,x_1y_m,\ldots,x_ny_1,\ldots,x_ny_m)$ an $nm$ dimensional
vector. For an $n\times m$ matrix $\bA=(\bA_{ij})$ and an $n'\times m'$ matrix
$\bB=(\bB_{kl})$ we construct similarly an $nn' \times mm'$ matrix
$\bA\otimes\bB = (\bA_{ij}\bB)$, i.e. each entry $\bA_{ij}$ in $\bA$ is
multiplied with a copy of the matrix or block $\bB$. That is: Given an $n
\times m$ matrix $\bA$ and a $k \times l$ matrix $\bB$ then $\bA \otimes \bB$
is the $nk \times ml$ matrix
\[
\bA \otimes \bB = 
\left(
\begin{array}{ccc}
a_{1,1}  & \ldots & a_{1,m} \\
\vdots & \ddots & \vdots \\
a_{n,1}  & \ldots & a_{m,n}
\end{array}
\right)
\otimes
\left(
\begin{array}{ccc}
b_{1,1}  & \ldots & b_{1,l} \\
\vdots & \ddots & \vdots \\
b_{k,1}  & \ldots & b_{k,l}
\end{array}
\right) =
\left(
\begin{array}{ccc}
a_{1,1}\bB & \ldots & a_{1,m}\bB \\
\vdots    & \ddots & \vdots \\
a_{n,1}\bB & \ldots & a_{n,m}\bB
\end{array}
\right)
\]

The tensor product of two vector spaces $\cV\otimes\cW$ can be defined as the
formal linear combinations of the tensor products $v_i \otimes w_j$ with $v_i$
and $w_j$ base vectors in $\cV$ and $\cW$, respectively.  For further details
we refer e.g to \cite[Chap.~14]{Roman05}.

Given a program $P$, our aim is to define compositionally a matrix
representing the program behaviour as a DTMC.  The domain of the associated
linear operator $\bT(P)$ is the space of {\em probabilistic configurations},
that is distributions over classical configurations, defined by
$\Dists(\Confs) = \Dists(\X^v \times \Labs) \subseteq \ell_2(\X^v\times
\Labs)$, where we identify a statement with its label, or more precisely, an
SOS configuration $\ip{S,s} \in \Confs$ with the pair $\ip{s,\init(S)} \in
\X^v \times \Labs$.

Among the building blocks of the construction of $\bT(P)$ are the {\em identity
  matrix} $\bI$ and the {\em matrix units} $\bE_{ij}$ containing only a single
non zero entry $(\bE_{ij})_{ij}=1$ and zero otherwise. We denote by $e_i$ the
unit vector with $(e_{i})_{i}=1$ and zero otherwise. As we represent
distributions by row vectors we use post-multiplication,
i.e. $\bT(x)=x\cdot\bT$.

A basic operator is the {\em update matrix} $\bU(c)$ which implements state
changes. The intention is that from an initial probabilistic state $\si$,
e.g. a distribution over classical states, we get a new probabilistic state
$\si'$ by the product $\si'=\si\cdot\bU$. The matrix $\bU(c)$ implements the
deterministic update of a variable to a constant $c$ via $(\bU(c))_{ij}=1$ if
$\xi(c)=j$ and $0$ otherwise, with $\xi:\X\rightarrow\N$ the underlying
enumeration of values in $\X$. In other words, this is a matrix which has only
one column (corresponding to $c$) containing $1$s while all other entries are
$0$.
Whatever the value of a variable is, after applying $\bU(c)$ to the state
vector describing the current situation we get a {\em point} distribution
expressing the fact that the value of our variable is now $c$.

We also define for any Boolean expression $b$ on $\X$ a diagonal {\em
  projection matrix} $\bP$ with $(\bP(b))_{ii} = 1$ if $b(c)$ holds and
$\xi(c)=i$ and $0$ otherwise.
The purpose of this diagonal matrix is to ``filter out'' only those states
which fulfil the condition $b$. If we want to apply an operator with matrix
representation $\bT$ only if a certain condition $b$ is fulfilled then
pre-multiplying this $\bP(b)\cdot\bT$ achieves this effect.

In Table~\ref{Matrices}\ we first define a multi-variable versions of the test
matrices and the update matrices via the tensor product `$\otimes$'. 



With the help of the auxiliary matrices we define for every program $P$ the
matrix $\bT(P)$ of the DTMC representing the program executions as the sum of
the effects of the individual control flow steps.  For each individual control
flow step it is of the form $\lsem{[B]^\ell} \otimes \bE_{\ell,\ell'}$ or
$\underline{\lsem{[B]^\ell}}\otimes\bE_{\ell,\ell'}$, where $(\ell,\ell')$ or
$(\ell,\underline\ell')\in\cF(P)$ and $\lsem{[B]^\ell}$ represents the
semantics of the block $B$ labelled by $\ell$.
The matrix $\bE_{\ell,\ell'}$ represents the control flow from label $\ell$ to
$\ell'$; it is a finite $l \times l$ matrix, where $l$ is the number of
(unique) distinct labels in $P$.

The definitions of $\lsem{[B]^\ell}$ and $\underline{\lsem{[B]^\ell}}$ are
given in Table~\ref{LOSop}. The semantics of an assignment block is
obviously given by $\bU(\varE{x}\gets e)$. For the random assignment 
we simply take the linear combination of assignments to all possible
values, weighted by the corresponding probability given by the
distribution $\rho$.
The semantics of a test block $[b]^\ell$ is given by its positive and its
negative part, both are test operators $\bP(b=\true)$ and $\bP(b=\false)$ 
as described before. 
The meaning of $\underline{\lsem{[B]^\ell}}$ is non-trivial only for tests $b$ 
while it is the identity for all the other blocks. The positive and negative 
semantics of all blocks is independent of the context and can be studied 
and  analysed in isolation from the rest of the program $P$.

\begin{table}[t]
\begin{center}
  \begin{minipage}{5cm}
    \[
    \begin{array}{rcl}
      \bP(s) & = & \displaystyle\bigotimes_{i=1}^{v} \bP(s(\varE{x}_i))
      \\[3mm]
      \bP(e = c) & = & \displaystyle\sum_{\cE(e)s=c} \bP(s)
    \end{array}
    \]
  \end{minipage}
  ~~~
  \begin{minipage}{5cm}
    \[
    \begin{array}{rcl}
      \bU(\varE{x}_k\gets c) & = & \displaystyle\bigotimes_{i=1}^{k-1}
      \bI
      \otimes
      \bU(c)
      \otimes
      \displaystyle\bigotimes_{i=k+1}^{v}
      \bI
      \\[3mm]
      \bU(\varE{x}_k\gets e) & = & \displaystyle\sum_{c} 
      \bP(e = c)
      \bU(\varE{x}_k\gets c)
    \end{array}
    \]
  \end{minipage}
  \end{center}
  \[
  \begin{array}{rclrcl}
    \lsem{[\getsS{x}{e}]^{\ell]}}
    & = & \bU(x \gets e) 
    &
    \lsem{[\ranS{v}{\rho}]^{\ell}}
    & = & \sum_{c\in\X} \rho(c)\bU(x \gets c) \\
    \lsem{[b]^{\ell}}
    & = & \bP(b=\false)
    &
    \underline{\lsem{[b]^{\ell}}}
    & = & \bP(b=\true)
    \\
    \multicolumn{6}{c}%
    {
      \lsem{[\skipS]^{\ell}}
      = 
      \underline{\lsem{[\skipS]^{\ell}}}
      =
      \underline{\lsem{[\getsS{x}{e}]^{\ell]}}}
      =
      \underline{\lsem{[\ranS{v}{\rho}]^{\ell}}}
      = \bI 
    } 
    \\
  \end{array}
  \]
  \caption{Elements of the LOS}
  \label{CollSem}
  \label{LOSop}
  \label{Matrices}
\end{table}

Based on the local (forward) semantics of each labelled block,
i.e. $\lsem{[B]^\ell}$ and $\underline{\lsem{[B]^\ell}}$, in $P$ we can define
the LOS semantics of $P$ as:
\[
\bT(P) = 
\sum_{(\ell,\ell')\in\cF(P)} 
     \lsem{[B]^\ell} \otimes \bE_{\ell,\ell'} +
\sum_{(\ell,\underline{\ell'})\in\cF(P)} 
     \underline{\lsem{[B]^\ell}}\otimes\bE_{\ell,\ell'}
\]

A minor adjustment is required to make our semantics conform to the DTMC
model.  As paths in a DTMC are \textit{maximal} (i.e. infinite) in the
underlying directed graph, we will add a single final loop via a virtual label
$\ell^*$. This corresponds to adding to
$\bT(P)$ the factor $\bI\otimes\bE_{\ell^*,\ell^*}$.

We first define a multi-variable versions of test matrices $\bP$ and update
matrices $\bU$ via the tensor product (see e.g.
\cite{Bertinoro10,APLAS13}). With the help of these auxiliary matrices we can
then define for every program $P$ the matrix $\bT(P)$ of the DTMC representing
the program executions as the sum of the effects of the individual control
flow steps.  For each individual control flow step it is of the form
$\lsem{[B]^\ell} \otimes \bE_{\ell,\ell'}$ or
$\underline{\lsem{[B]^\ell}}\otimes\bE_{\ell,\ell'}$, where $(\ell,\ell')$ or
$(\ell,\underline\ell')\in\cF(P)$ and $\lsem{[B]^\ell}$ represents the
semantics of the block $B$ labelled by $\ell$. The matrix $\bE_{\ell,\ell'}$
represents the control flow from label $\ell$ to $\ell'$; it is a finite $l
\times l$ matrix, where $l$ is the number of (unique) distinct labels in $P$.
The definitions of $\lsem{[B]^\ell}$ and $\underline{\lsem{[B]^\ell}}$ are
given in Table~\ref{LOSop}.
%
%
Based on the local semantics of each labelled block, i.e. $\lsem{[B]^\ell}$
and $\underline{\lsem{[B]^\ell}}$, in $P$ we can define the LOS semantics of
$P$ as:
\[
\bT(P) = 
\sum_{(\ell,\ell')\in\cF(P)} 
     \lsem{[B]^\ell} \otimes \bE_{\ell,\ell'} +
\sum_{(\ell,\underline{\ell'})\in\cF(P)} 
     \underline{\lsem{[B]^\ell}}\otimes\bE_{\ell,\ell'}
\]

A minor adjustment is required to make our semantics conform to the DTMC
model.  As paths in a DTMC are \textit{maximal} (i.e. infinite) in the
underlying directed graph, we will add a single final loop via a virtual
label $\ell^*$ as discussed in Section~\ref{Language}. This corresponds to
adding to $\bT(P)$ the factor $\bI\otimes\bE_{\ell^*,\ell^*}$.


\subsection{LOS and PAI}

Important for the applicability of PAI in the context of program analysis is
the fact that it possesses nice compositionality properties. These allows us
to construct the abstract semantics by abstracting the single blocks of the
concrete semantics $\bT(S)$:
\[
\bT(S)^\# = 
\bA^\dagger \bT(S) \bA =
\bA^\dagger \left( \sum\lsem{[B]^\ell} \right) \bA =
\sum \left( \bA^\dagger \lsem{[B]^\ell} \bA \right) =
\sum \lsem{[B]^\ell}^\#
\]
(where, for simplicity, we do not distinguish between the positive and
negative semantics of blocks). The fact that we can work with the abstract
semantics of individual blocks instead of the full operator obviously reduces
the complexity of the analysis substantially.  Another important fact is that
the Moore-Penrose pseudo-inverse of a tensor product can be computed as
\cite[2.1,Ex~3]{BenIsraelGreville03}:
\[
(\bA_1 \otimes \bA_2 \otimes \ldots \otimes \bA_v)^\dagger
=
\bA_1^\dagger \otimes \bA_2^\dagger \otimes \ldots \otimes \bA_v^\dagger.
\]
We can therefore abstract properties of individual variables and then combine
them in the global abstraction. This is also made possible by the definition
of the concrete LOS semantics which is heavily based on the use of tensor
product. Typically we have $\lsem{[B]^\ell} = \left( \bigotimes_{i=1}^v
  \bT_{i\ell} \right) \otimes \bE_{\ell\ell'}$ or a sum of a few of such
terms. The $\bT_{i\ell}$ represents the effect of $\bT(S)$, and in particular
of $\lsem{[B]^\ell}$, on variable $i$ at label $\ell$ (both labels and
variables only form a finite set). For example, we can define an abstraction
$\bA$ for one variable and apply it individually to all variables
(e.g. extracting their even/odd property), or use different abstractions for
different variables (maybe even forgetting about some of them by using
$\bA_f=(1,1,\ldots)^t$) and define $\bA=\bigotimes_{i=1}^v \bA_i$ such that
$\bA^\dagger=\bigotimes_{i=1}^v \bA_i^\dagger$ in order to get an analysis on
the full state space.


\subsection{Relation to Other Probabilistic Semantics}

We can use the LOS to reconstruct the semantics of Kozen \cite{Kozen81}\ by
simply taking the limit of $\bT(S)^n(s_0)$ for $n\rightarrow\infty$ for all
initial states $s_0$. The limit state $\bT(S)^n(s_0)$ still contains too much
information in relation to Kozen's semantics; in fact we only need the
probability distributions on the possible values of the variables at the final
label.

In order to extract information about the probability that variables have
certain values at a certain label, i.e. program point $\ell$, we can use the
operator $\bI\otimes \ldots \otimes \bI \otimes \bE_{\ell,\ell}$. In
particular, for extracting the information about a probabilistic state 
we will 
use
$
\bS_\ell = \bI \otimes \ldots \otimes \bI \otimes e_\ell
$ 
which forgets about all distributions at other labels than $\ell$. In
particular we use $\bS_f$ for the final looping $\stopS$ statement and $\bS_i$
for the initial label $\init(P)$ of the program. We denote by $e_0$ the
base vector in $\R^l$ which expresses the fact that we are in the initial label,
i.e. $e_0=e_{\init(P)}$.

\begin{proposition}[\cite{APLAS13}]
  Given a program $P$ and an initial state $s_0$ as a distribution over the
  program variables, then $(s_0 \otimes e_0) \bT(P)^n \bS_f$ corresponds to
  the distribution over all states on which $P$ terminates in $n$ or less
  computational steps.
\end{proposition}

We can now show that the effect of the LOS operator we obtain as solution to
Kozen's fixed-point equations agrees with the ``output''
$\lim_{n\rightarrow\infty} (s_0\otimes e_0)\bT^n\bS_f$ we get via the
LOS. Essentially, both semantics define the same I/O operator, provided we
supply them with the appropriate input.
However, the LOS also provides information about internal labels and reflects
the relation between different variables via the tensor product representation.

\begin{proposition}[\cite{APLAS13}]
  Given a program $P$ and an initial probabilistic state $s_0$ as a
  distribution over the program variables, let $\ksem{P}$ be Kozen's semantics
  of $P$ and $\bT(P)$ the LOS. Then
  $
  (s_0 \otimes e_0) (\lim_{n\rightarrow\infty} \bT^n) \bS_f = 
    s_0 \ksem{P}.
  $
\end{proposition}


\section{An Example: Monty Hall}
\label{Monty}

The origins of this example are legendary. Allegedly, it goes back to some TV
show in which the contestant was given the chance to win a car or other prizes
by picking the right door behind which the desired prize could be found.

The game proceeds as follows: First the contestant is invited to pick one of
three doors (behind one is the prize) but the door is not yet opened. Instead,
the host -- legendary Monty Hall -- opens one of the other doors which is
empty. After that the contestant is given a last chance to stick with his/her
door or to switch to the other closed one. Note that the host (knowing where
the prize is) has always at least one door he can open.

The problem is whether it is better to stay stubborn or to switch the chosen
door.  Assuming that there is an equal chance for all doors to hide the prize
it is a favourite exercise in basic probability theory to demonstrate that it
is better to switch to a new door.

We will analyse this example using probabilistic techniques in program
analysis - rather than more or less informal mathematical arguments. An
extensive discussion of the problem can be found in \cite{Stirzaker99}\ where
it is also observed that a bias in hiding the car (e.g. because the
architecture of the TV studio does not allow for enough room behind a door to
put the prize there) changes the analysis dramatically. 


We first consider two programs $H_t$ and $H_w$ in which the contestant is
either sticking to his/her initial choice or where there is a switch to (the
other closed) door:
\begin{center}
\begin{minipage}[t]{7cm}
\begin{verbatim}
# Pick winning door
d ?= {0,1,2};
# Pick guess
g ?= {0,1,2};
# Open empty door
o ?= {0,1,2};
while ((o == g) || (o == d)) do
  o := (o+1)%3;
od;
\end{verbatim}
\end{minipage}
\begin{minipage}[t]{7cm}
\begin{verbatim}
# Pick winning door
d ?= {0,1,2};
# Pick guess
g ?= {0,1,2};
# Open empty door
o ?= {0,1,2};
while ((o == g) || (o == d)) do
  o := (o+1)%3;
od;
# Switch
g := (g+1)%3;
while (g == o) do
  g := (g+1)%3;
od;
\end{verbatim}
\end{minipage}
\end{center}

We introduce a short hand notation (macros) with $\ell$ any program label:
\begin{eqnarray*}
\bT_{\mbox{pick}} 
&=&
\frac{1}{3} \left(
\bU({\tt d} \gets 0) +
\bU({\tt d} \gets 1) +
\bU({\tt d} \gets 2)
\right) 
\otimes \bE(1,2) + \\
& &
\frac{1}{3} \left(
\bU({\tt g} \gets 0) +
\bU({\tt g} \gets 1) +
\bU({\tt g} \gets 2)
\right) 
\otimes \bE(2,3) + \\
& &
\frac{1}{3} \left(
\bU({\tt o} \gets 0) +
\bU({\tt o} \gets 1) +
\bU({\tt o} \gets 2)
\right) 
\otimes \bE(3,4) + \\
& & \bP({\tt (o==g)||(o==d)}=\true)  \otimes \bE(4,5) + \\ & & 
    \bP({\tt (o==g)||(o==d)}=\false) \otimes \bE(4,6) + \\ 
& & 
\bU({\tt o} \gets (o+1)\%3) \otimes \bE(5,4)
\end{eqnarray*}
and (parametric in the initial label of the final part of $H_w$):
\begin{eqnarray*}
\bT_{\mbox{flip}}(\ell)
&=& \bU({\tt g} \gets (g+1)\%3) \otimes \bE(\ell,\ell+1) + \\
& & \bP({\tt (g==o)}=\true)  \otimes \bE(\ell+1,\ell+2) + \\ & & 
    \bP({\tt (g==o)}=\false) \otimes \bE(\ell+1,\ell+3) + \\
& & \bU({\tt g} \gets (g+1)\%3) \otimes \bE(\ell,\ell+1)
\end{eqnarray*} 
The LOS semantics of the two programs can then be easily specified:
\begin{eqnarray*}
\bT(H_t) 
&=& \bT_{\mbox{pick}} + \bI \otimes \bE(6,6) \\
\bT(H_w)
&=& \bT_{\mbox{pick}} + \bT_{\mbox{flip}}(6) + \bI \otimes \bE(9,9)
\end{eqnarray*}

The matrix representations of the individual transfer operators $\bP$, $\bU$
etc. and the complete LOS semantics of both programs (as $162\times162$ or
$243\times243$ matrices) are given in detail in \cite{Bertinoro10}. We can
compute these matrices via an experimental tool ``{\tt pwc}'' which has been
written in {\tt OCaml} and which generates the matrices to be used with the
numerical tool {\tt octave}. These matrices are based on the enumeration of
elements in $\States$ as follows:
\begin{small}
\[
\begin{array}{rcl}
 1 & \ldots & ({\tt d} \mapsto 0, {\tt g} \mapsto 0, {\tt o} \mapsto 0) \\
 2 & \ldots & ({\tt d} \mapsto 0, {\tt g} \mapsto 0, {\tt o} \mapsto 1) \\
 3 & \ldots & ({\tt d} \mapsto 0, {\tt g} \mapsto 0, {\tt o} \mapsto 2) \\
 4 & \ldots & ({\tt d} \mapsto 0, {\tt g} \mapsto 1, {\tt o} \mapsto 0) \\
 5 & \ldots & ({\tt d} \mapsto 0, {\tt g} \mapsto 1, {\tt o} \mapsto 1) \\
 6 & \ldots & ({\tt d} \mapsto 0, {\tt g} \mapsto 1, {\tt o} \mapsto 2) \\
 7 & \ldots & ({\tt d} \mapsto 0, {\tt g} \mapsto 2, {\tt o} \mapsto 0) \\
 8 & \ldots & ({\tt d} \mapsto 0, {\tt g} \mapsto 2, {\tt o} \mapsto 1) \\
 9 & \ldots & ({\tt d} \mapsto 0, {\tt g} \mapsto 2, {\tt o} \mapsto 2) \\
\end{array}
~~
\begin{array}{rcl}
10 & \ldots & ({\tt d} \mapsto 1, {\tt g} \mapsto 0, {\tt o} \mapsto 0) \\
11 & \ldots & ({\tt d} \mapsto 1, {\tt g} \mapsto 0, {\tt o} \mapsto 1) \\
12 & \ldots & ({\tt d} \mapsto 1, {\tt g} \mapsto 0, {\tt o} \mapsto 2) \\
13 & \ldots & ({\tt d} \mapsto 1, {\tt g} \mapsto 1, {\tt o} \mapsto 0) \\
14 & \ldots & ({\tt d} \mapsto 1, {\tt g} \mapsto 1, {\tt o} \mapsto 1) \\
15 & \ldots & ({\tt d} \mapsto 1, {\tt g} \mapsto 1, {\tt o} \mapsto 2) \\
16 & \ldots & ({\tt d} \mapsto 1, {\tt g} \mapsto 2, {\tt o} \mapsto 0) \\
17 & \ldots & ({\tt d} \mapsto 1, {\tt g} \mapsto 2, {\tt o} \mapsto 1) \\
18 & \ldots & ({\tt d} \mapsto 1, {\tt g} \mapsto 2, {\tt o} \mapsto 2) \\
\end{array}
~~
\begin{array}{rcl}
19 & \ldots & ({\tt d} \mapsto 2, {\tt g} \mapsto 0, {\tt o} \mapsto 0) \\
20 & \ldots & ({\tt d} \mapsto 2, {\tt g} \mapsto 0, {\tt o} \mapsto 1) \\
21 & \ldots & ({\tt d} \mapsto 2, {\tt g} \mapsto 0, {\tt o} \mapsto 2) \\
22 & \ldots & ({\tt d} \mapsto 2, {\tt g} \mapsto 1, {\tt o} \mapsto 0) \\
23 & \ldots & ({\tt d} \mapsto 2, {\tt g} \mapsto 1, {\tt o} \mapsto 1) \\
24 & \ldots & ({\tt d} \mapsto 2, {\tt g} \mapsto 1, {\tt o} \mapsto 2) \\
25 & \ldots & ({\tt d} \mapsto 2, {\tt g} \mapsto 2, {\tt o} \mapsto 0) \\
26 & \ldots & ({\tt d} \mapsto 2, {\tt g} \mapsto 2, {\tt o} \mapsto 1) \\
27 & \ldots & ({\tt d} \mapsto 2, {\tt g} \mapsto 2, {\tt o} \mapsto 2) \\
\end{array}
\]
\end{small}


We can use the LOS semantics to analyse whether it is $H_t$ or $H_w$ that
implements the better strategy. In principle, we can do this using the
concrete semantics we constructed above. However, it is rather cumbersome to
work with ``relatively large'' $162\times162$ or $243\times243$ matrices, even
when they are sparse, i.e. contain almost only zeros (in fact only about
$1.2\%$ of entries in $H_t$ and $0.7\%$ of entries in $H_w$ are non-zero).

If we want to analyse the final states, i.e. which of the two programs has a
better chance of getting the right door, we need to start with an initial
configuration and then iterate $\bT(H)$ until we reach a/the final
configuration. For our programs it is sufficient to indicate that we start in
label $1$, while the state is irrelevant as we initialise all three variables
at the beginning of the program; we could take -- for example -- a state with
$d=o=g=0$. The vector or distribution which describes this initial
configuration is a $162$ or $243$ dimensional vector. We can describe it in a
rather compact form as:
\[
  \vec{x}_0 =
  \left(
    \begin{array}{ccc}
      1 & 0 & 0
    \end{array}
  \right)
  \otimes
  \left(
    \begin{array}{ccc}
      1 & 0 & 0
    \end{array}
  \right)
  \otimes
  \left(
    \begin{array}{ccc}
       1 & 0 & 0
   \end{array}
  \right)
  \otimes
  \left(
    \begin{array}{ccccc}
       1 & 0 & 0 & \ldots & 0
   \end{array}
  \right),
\]
where the last factor is $6$ or $9$ dimensional, depending on whether we deal
with $H_t$ or $H_w$. This represents a point distribution on $162$ or $243$
relevant distributions.

Assuming that our program terminates for all initial states, as it is the case
here, then there exists a certain number of iterations $t$ such that
$\vec{x}_0\bT(H)^t = \vec{x}_0\bT(H)^{t+1}$, i.e. we will eventually reach a
fix-point which gives us a distribution over configurations. In general, as in
our case here, this will not be just a point distribution. Again we get
vectors of dimension $162$ or $243$, respectively. For $H_t$ and $H_w$ there
are $12$ configurations which have non-zero probability.
  \[
  \mbox{for}~H_t~~
  \left\{
  \begin{array}{rcl}
  x_{12} &=&  0.074074 \\
  x_{18} &=&  0.037037 \\
  x_{36} &=&  0.11111 \\
  x_{48} &=&  0.11111 \\
  x_{72} &=&  0.11111 \\
  x_{78} &=&  0.037037 \\
  x_{90} &=&  0.074074 \\
  x_{96} &=&  0.11111 \\
  x_{120} &=&  0.11111 \\
  x_{132} &=&  0.11111 \\
  x_{150} &=&  0.074074 \\
  x_{156} &=&  0.037037
  \end{array}
  \right.
  ~~~~~~~
  \mbox{for}~H_w~~
 \left\{
  \begin{array}{rcl}
  x_{18} &=&  0.11111  \\
  x_{27} &=&  0.11111 \\
  x_{54} &=&  0.037037 \\
  x_{72} &=&  0.074074 \\
  x_{108} &=&  0.074074 \\
  x_{117} &=&  0.11111 \\
  x_{135} &=&  0.11111 \\
  x_{144} &=&  0.037037 \\
  x_{180} &=&  0.037037 \\
  x_{198} &=&  0.074074 \\
  x_{225} &=&  0.11111 \\
  x_{234} &=&  0.11111
  \end{array}
  \right.
\]

It is anything but easy to determine from this information which of the two
strategies is more successful. In order to achieve this we will abstract away
all unnecessary information. First, we ignore the syntactic information: If we
are in the terminal state, then we have reached the final {\tt stop} state,
but even if this would not be the case we only need to know whether in the
final state we have guessed the right door, i.e. whether {\tt d==g} or not. We
thus also don't need to know the value of {\tt o} as it ultimately is of no
interest to us which door had been opened during the game. Therefore, we can
use the forgetful abstraction $\bA_f$ to simplify the information contained in
the terminal state. Regarding {\tt d} and {\tt g} we want to know everything,
and thus use the trivial abstraction $\bA=\bI$, i.e. the identity. The result
for $H_t$ is for $\vec{x}_t$ the terminal configuration distribution as well
as for $H_w$ with terminal distribution $\vec{x}_w$
\begin{eqnarray*}
\vec{x}_t\cdot(\bI\otimes\bI\otimes\bA_f\otimes\bA_f)
& = &
\left(
  \begin{array}{rrrrrrrrr}
    0.11 & 0.11 & 0.11 & 0.11 & 0.11 & 0.11 & 0.11 & 0.11 & 0.11 \\
  \end{array}
\right)
\\
\vec{x}_w\cdot(\bI\otimes\bI\otimes\bA_f\otimes\bA_f)
& = &
\left(
  \begin{array}{rrrrrrrrr}
    0.22 & 0.04 & 0.07 & 0.07 & 0.22 & 0.04 & 0.04 & 0.07 & 0.22 \\
  \end{array}
\right)
\end{eqnarray*}
The nine coordinates of these vectors correspond to 
  $({\tt d}\mapsto0, {\tt g} \mapsto0)$, 
  $({\tt d}\mapsto0, {\tt g} \mapsto1)$, 
  $({\tt d}\mapsto0, {\tt g} \mapsto2)$, 
  $({\tt d}\mapsto1, {\tt g} \mapsto0)$, \ldots,
  $({\tt d}\mapsto2, {\tt g} \mapsto2)$.
This is in principle enough to conclude that $H_w$ is the better strategy.

However, we can go a step further and abstract not the values of {\tt d} and
{\tt g} but their relation, i.e. whether they are equal or different. For this
we need the abstraction:
\[
  \bA_w = 
  \left(
    \begin{array}{rr}
      1 & 0 \\
      0 & 1 \\
      0 & 1 \\
      0 & 1 \\
      1 & 0 \\
      0 & 1 \\
      0 & 1 \\
      0 & 1 \\
      1 & 0 \\
    \end{array}
  \right)
\]
where the first column corresponds to a winning situation (i.e.  {\tt d} and
{\tt g} are equal), and the second to unequal {\tt d} and {\tt g}. With this
we get for $H_t$ and $H_w$ respectively:
\begin{eqnarray*}
\vec{x}\cdot(\bA_w\otimes\bA_f\otimes\bA_f)
& = &
\left(
  \begin{array}{rr}
    0.33333 & 0.66667 \\
  \end{array}
\right)
\\
\vec{x}\cdot(\bA_w\otimes\bA_f\otimes\bA_f)
& = &
\left(
  \begin{array}{rr}
    0.66667 & 0.33333 \\
  \end{array}
\right)
\end{eqnarray*}
It is now obvious that $H_t$ has just a $\frac{1}{3}$ chance of winning, while
$H_w$ has a $\frac{2}{3}$ probability of picking the winning door.


We can also consider a more general strategy which is a combination of {\em
  switching} and {\em sticking}: With a certain probability $p$ the contestant
switches or sticks with his/her first choice. This corresponds to the
following program(s) $H(p)$ parametrised using the probability $p\in[0,1]$:
\begin{center}
\begin{minipage}[t]{8cm}
\begin{verbatim}
# Pick winning door
d ?= {0,1,2};
# Pick guess
g ?= {0,1,2};
# Open empty door
o ?= {0,1,2};
while ((o == g) || (o == d)) do
  o := (o+1)%3;
od;
# Switch or stick with probability p
choose p: g := (g+1)%3;
          while (g == o) do
          g := (g+1)%3; od
or (1-p): skip
ro;
\end{verbatim}
\end{minipage}
\end{center}

One can now analyse the winning chance of these programs $H(p)$ depending on
the parameter $p$ and see what chance of winning we have. The LOS is given by
\begin{eqnarray*}
\bT(H(p))
&=& \bT_{\mbox{pick}} + 
    p\cdot\bI\otimes \bE(6,7) + (1-p)\cdot\bI\otimes\bE(6,10) 
+ \bT_{\mbox{flip}}(7) + \bI\otimes\bE(10,11) + \bI\otimes\bE(11,11)
\end{eqnarray*}
One way to interpret this is to see this as the problem of determining the
optimal strategy by identifying the optimal value for $p$ for which the
winning chance is the largest, i.e. as an optimisation problem with objective
function:
\[
\Phi(p) = s_0 \cdot (\lim_{n\rightarrow\infty} \bT(H(p))^n) \cdot 
          (\bA_w\otimes\bA_f\otimes\bA_f) \cdot \bP_1
\]
where $\bT(H(p))$ is the LOS semantics of $H(p)$ for a certain value of $p$
and $\bP_1$ is the projection of the first coordinate (in $\R^2$). In effect
we can avoid the limit as our program always terminates after a finite number
of steps; it is also independent of the initial state $s_0$. This way we have
the optimisation problem:
\[
\max \Phi(p) 
~~\mbox{subject to}~~~
0 \leq p \leq 1
\]

The relationship between the probability or parameter $p$ and the chances of
winning (obtained using {\tt octave} for various values of $p$) can be seen in
the following diagram.

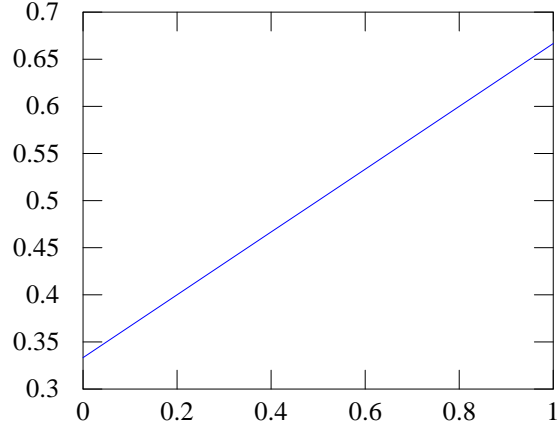
\begin{center}
\begin{tikzpicture}[gnuplot]
\path (0.000,0.000) rectangle (8.000,6.000);
\gpcolor{color=gp lt color border}
\gpsetlinetype{gp lt border}
\gpsetlinewidth{0.50}
\draw[gp path] (1.196,0.616)--(1.447,0.616);
\draw[gp path] (7.447,0.616)--(7.196,0.616);
\gpcolor{rgb color={0.000,0.000,0.000}}
\node[gp node right,font={\fontsize{10pt}{12pt}\selectfont}] at (1.012,0.616) {0.3};
\gpcolor{color=gp lt color border}
\draw[gp path] (1.196,1.243)--(1.447,1.243);
\draw[gp path] (7.447,1.243)--(7.196,1.243);
\gpcolor{rgb color={0.000,0.000,0.000}}
\node[gp node right,font={\fontsize{10pt}{12pt}\selectfont}] at (1.012,1.243) {0.35};
\gpcolor{color=gp lt color border}
\draw[gp path] (1.196,1.870)--(1.447,1.870);
\draw[gp path] (7.447,1.870)--(7.196,1.870);
\gpcolor{rgb color={0.000,0.000,0.000}}
\node[gp node right,font={\fontsize{10pt}{12pt}\selectfont}] at (1.012,1.870) {0.4};
\gpcolor{color=gp lt color border}
\draw[gp path] (1.196,2.497)--(1.447,2.497);
\draw[gp path] (7.447,2.497)--(7.196,2.497);
\gpcolor{rgb color={0.000,0.000,0.000}}
\node[gp node right,font={\fontsize{10pt}{12pt}\selectfont}] at (1.012,2.497) {0.45};
\gpcolor{color=gp lt color border}
\draw[gp path] (1.196,3.123)--(1.447,3.123);
\draw[gp path] (7.447,3.123)--(7.196,3.123);
\gpcolor{rgb color={0.000,0.000,0.000}}
\node[gp node right,font={\fontsize{10pt}{12pt}\selectfont}] at (1.012,3.123) {0.5};
\gpcolor{color=gp lt color border}
\draw[gp path] (1.196,3.750)--(1.447,3.750);
\draw[gp path] (7.447,3.750)--(7.196,3.750);
\gpcolor{rgb color={0.000,0.000,0.000}}
\node[gp node right,font={\fontsize{10pt}{12pt}\selectfont}] at (1.012,3.750) {0.55};
\gpcolor{color=gp lt color border}
\draw[gp path] (1.196,4.377)--(1.447,4.377);
\draw[gp path] (7.447,4.377)--(7.196,4.377);
\gpcolor{rgb color={0.000,0.000,0.000}}
\node[gp node right,font={\fontsize{10pt}{12pt}\selectfont}] at (1.012,4.377) {0.6};
\gpcolor{color=gp lt color border}
\draw[gp path] (1.196,5.004)--(1.447,5.004);
\draw[gp path] (7.447,5.004)--(7.196,5.004);
\gpcolor{rgb color={0.000,0.000,0.000}}
\node[gp node right,font={\fontsize{10pt}{12pt}\selectfont}] at (1.012,5.004) {0.65};
\gpcolor{color=gp lt color border}
\draw[gp path] (1.196,5.631)--(1.447,5.631);
\draw[gp path] (7.447,5.631)--(7.196,5.631);
\gpcolor{rgb color={0.000,0.000,0.000}}
\node[gp node right,font={\fontsize{10pt}{12pt}\selectfont}] at (1.012,5.631) {0.7};
\gpcolor{color=gp lt color border}
\draw[gp path] (1.196,0.616)--(1.196,0.867);
\draw[gp path] (1.196,5.631)--(1.196,5.380);
\gpcolor{rgb color={0.000,0.000,0.000}}
\node[gp node center,font={\fontsize{10pt}{12pt}\selectfont}] at (1.196,0.308) {0};
\gpcolor{color=gp lt color border}
\draw[gp path] (2.446,0.616)--(2.446,0.867);
\draw[gp path] (2.446,5.631)--(2.446,5.380);
\gpcolor{rgb color={0.000,0.000,0.000}}
\node[gp node center,font={\fontsize{10pt}{12pt}\selectfont}] at (2.446,0.308) {0.2};
\gpcolor{color=gp lt color border}
\draw[gp path] (3.696,0.616)--(3.696,0.867);
\draw[gp path] (3.696,5.631)--(3.696,5.380);
\gpcolor{rgb color={0.000,0.000,0.000}}
\node[gp node center,font={\fontsize{10pt}{12pt}\selectfont}] at (3.696,0.308) {0.4};
\gpcolor{color=gp lt color border}
\draw[gp path] (4.947,0.616)--(4.947,0.867);
\draw[gp path] (4.947,5.631)--(4.947,5.380);
\gpcolor{rgb color={0.000,0.000,0.000}}
\node[gp node center,font={\fontsize{10pt}{12pt}\selectfont}] at (4.947,0.308) {0.6};
\gpcolor{color=gp lt color border}
\draw[gp path] (6.197,0.616)--(6.197,0.867);
\draw[gp path] (6.197,5.631)--(6.197,5.380);
\gpcolor{rgb color={0.000,0.000,0.000}}
\node[gp node center,font={\fontsize{10pt}{12pt}\selectfont}] at (6.197,0.308) {0.8};
\gpcolor{color=gp lt color border}
\draw[gp path] (7.447,0.616)--(7.447,0.867);
\draw[gp path] (7.447,5.631)--(7.447,5.380);
\gpcolor{rgb color={0.000,0.000,0.000}}
\node[gp node center,font={\fontsize{10pt}{12pt}\selectfont}] at (7.447,0.308) {1};
\gpcolor{color=gp lt color border}
\draw[gp path] (1.196,5.631)--(1.196,0.616)--(7.447,0.616)--(7.447,5.631)--cycle;
\gpcolor{rgb color={0.000,0.000,1.000}}
\gpsetlinetype{gp lt plot 0}
\draw[gp path] (1.196,1.034)--(1.821,1.452)--(2.446,1.870)--(3.071,2.288)--(3.696,2.706)%
  --(4.322,3.123)--(4.947,3.541)--(5.572,3.959)--(6.197,4.377)--(6.822,4.795)--(7.447,5.213);
\gpdefrectangularnode{gp plot 1}{\pgfpoint{1.196cm}{0.616cm}}{\pgfpoint{7.447cm}{5.631cm}}
\end{tikzpicture}

\end{center}

As one would expect the worst chance of winning is for $p=0$, i.e. for the
{\em sticking} strategy, and the best (of $\frac{2}{3}$) for $p=1$, i.e. {\em
  switching}. This is a very simple case of a synthesis problem, but it
illustrated the basic idea: We optimise a continuous parameter $p$ in order
to get a program with optimal performance (winning chance).


\section{An Example: Swapping Variables}
\label{Results}

We consider another simple situation to illustrate how non-linear optimisation
can be used to general or transform programs such that certain requirements
are fulfilled.


Given a number of basic blocks we aim in constructing a (small) program which
exchanges two variables $x$ and $y$. We assume -- to keep the setting as
simple as possible -- that $x$ and $y$ can only take two values $0$ and $1$.

If we consider the state space for these two variables we need to consider the
tensor product $\cV(\{0,1\}\times\{0,1\}) = \cV(\{0,1\}) \otimes \cV(\{0,1\})
= \R^2\otimes\R^2 = \R^4$. In this four dimensional space the first dimension
corresponds to the (classical) state $s_1=[x\mapsto0,y\mapsto0]$, the second
one to $s_2=[x\mapsto0,y\mapsto0]$, the third to $s_3=[x\mapsto1,y\mapsto0]$,
and the forth to $s_4=[x\mapsto1,y\mapsto1]$.

The swapping operation we aim to implement is thus represented by the matrix
\[
\bS=
\left(
\begin{array}{cccc}
1 & 0 & 0 & 0 \\
0 & 0 & 1 & 0 \\
0 & 1 & 0 & 0 \\
0 & 0 & 0 & 1 \\
\end{array}
\right).
\]
If the values of $x$ and $y$ are the same nothing happens when we swap them,
thus the entry $1$ in the diagonal corresponding to the first and forth
classical state. For the second and third coordinate we only have to exchange
the probabilities associated to these two classical states.


We consider a few basic building blocks with which we aim to achieve the task
of implementing this simple specification. We try to have several options to
achieve our aim and thus allow also for a buffer variable $z$ which we might
use to swap $x$ and $y$. Another well known way is to use the `exclusive or'
(xor) to swap $x$ and $y$. In our case we implement xor as $x \oplus y = (x+y)
\bmod 2$.









What we aim for is a program $P$ of the form (allowing in the obvious way for
an n-array choice):
\[
\begin{array}{l}
\mbox{\tt choose}~\la_{1,1}:S_{1}~\mbox{\tt or}~
                  \la_{1,2}:S_{2}~\mbox{\tt or}~\ldots 
   ~\mbox{\tt or}~\la_{1,13}:S_{23} ~{\tt ro}; \\
\mbox{\tt choose}~\la_{2,1}:S_{1}~\mbox{\tt or}~
                  \la_{2,2}:S_{2}~\mbox{\tt or}~\ldots 
   ~\mbox{\tt or}~\la_{2,13}:S_{13} ~{\tt ro}; \\
\mbox{\tt choose}~\la_{3,1}:S_{1}~\mbox{\tt or}~
                  \la_{3,2}:S_{2}~\mbox{\tt or}~\ldots 
   ~\mbox{\tt or}~\la_{3,13}:S_{13} ~{\tt ro};\\
\end{array}
\]
where we have $13$ different elementary blocks $S_{i,j}$ which we enumerate as
follows:
\[
\begin{array}{c}
     \skipL{1}       
~~~  \getsL{x}{y}{2} 
~~~  \getsL{x}{z}{3} 
~~~  \getsL{y}{x}{4} 
~~~  \getsL{y}{z}{5} 
~~~  \getsL{z}{x}{6} 
~~~  \getsL{z}{y}{7} 
  \\
     \getsL{x}{(x+y) \bmod 2}{8} 
~~~  \getsL{x}{(x+z) \bmod 2}{9} 
~~~  \getsL{y}{(y+x) \bmod 2}{10} 
  \\
     \getsL{y}{(y+z) \bmod 2}{11} 
~~~  \getsL{z}{(z+x) \bmod 2}{12} 
~~~  \getsL{z}{(z+y) \bmod 2}{13} 
\end{array}
\]
such that $\lim_{t\rightarrow\infty}\bT(P)^t=\bS\otimes\bE_{i,f}$, i.e. the
program does in three steps what we expect from $\bS$ (and control transfers
from the initial label $i$ to the final $f$). We ignore the control flow, we
are just interested in the transfer functions associated to the basic blocks.

The LOS program semantics of the program we aim in generating is made up from
this $13$ transfer functions $\bF_1\ldots\bF_{13}$ with $\bF_{j}=\lsem{S_j}$, i.e.
\[
\bT = \sum_{i=1}^3 \bT_i
~~~\mbox{with}~~~
\bT_i = \sum_{j=1}^{13} \la_{ij} \bF_j
\]

Each of the $\bF_i$ are constructed as $8\times8$ matrices on the tensor
product space $\cV(\{0,1\}\times\{0,1\}\times\{0,1\}) = \cV(\{0,1\}) \otimes
\cV(\{0,1\}) \otimes \cV(\{0,1\}) = \R^2\otimes\R^2\otimes\R^2 = \R^8$.  In
this eight dimensional space the dimensions corresponds to the (classical)
states:
\begin{center}
\begin{minipage}{7.5cm}
\begin{eqnarray*}
s_1 & \ldots & [x\mapsto0,y\mapsto0,z\mapsto0] \\
s_2 & \ldots & [x\mapsto0,y\mapsto0,z\mapsto1] \\
s_3 & \ldots & [x\mapsto0,y\mapsto1,z\mapsto0] \\
s_4 & \ldots & [x\mapsto0,y\mapsto1,z\mapsto1] \\
\end{eqnarray*}
\end{minipage}
\begin{minipage}{7.5cm}
\begin{eqnarray*}
s_5 & \ldots & [x\mapsto1,y\mapsto0,z\mapsto0] \\
s_6 & \ldots & [x\mapsto1,y\mapsto0,z\mapsto1] \\
s_7 & \ldots & [x\mapsto1,y\mapsto1,z\mapsto0] \\
s_8 & \ldots & [x\mapsto1,y\mapsto1,z\mapsto1] \\
\end{eqnarray*}
\end{minipage}
\end{center}

As we do not care what value $z$ has in the end we can abstract it away using
a technique called Probabilistic Abstract Interpretation (PAI) using the
abstraction operator $\bA=\bI\otimes\bI\otimes\bA_f$ (with $\bA_f$ the
forgetful abstraction) and its concretisation $\bG$ given by the Moore-Penrose
pseudo-inverse. 
\[
\bA = 
\left(
\begin{array}{cccc}
1 & 0 & 0 & 0 \\
0 & 1 & 0 & 0 \\
0 & 0 & 1 & 0 \\
0 & 0 & 0 & 1 \\
\end{array}
\right)
\otimes
\left(
\begin{array}{c}
1 \\
1 \\
\end{array}
\right)
\]
and its concretisation function given by the Moore-Penrose pseudo-inverse:
\[
\bG =
\bA^\dagger = 
\left(
\begin{array}{cccc}
1 & 0 & 0 & 0 \\
0 & 1 & 0 & 0 \\
0 & 0 & 1 & 0 \\
0 & 0 & 0 & 1 \\
\end{array}
\right)
\otimes
\left(
\begin{array}{cc}
\frac{1}{2} & \frac{1}{2} \
\end{array}
\right)
\]

With this our main objective function, describing the requirement that we want
a program $\bT(\la_{ij})$ which implements the swap of $x$ and $y$, is given
by:
\[
\Phi_{00}(\lambda_{ij}) =
\| \bA^\dagger\bT(\lambda_{ij})\bA - \bS \|_2
\]
We also use a  general objective function which penalises for reading
or writing to the third variable $z$:
\[
\Phi_{\rho\omega}(\lambda_{ij}) =
\| \bA^\dagger\bT(\lambda_{ij})\bA - \bS \|_2 + 
   \rho R(\lambda_{ij}) + \omega W(\lambda_{ij})
\]
where the function $R$ and $W$ determine the probability that in each step of
our program the variable $z$ is read or written to respectively. Define two
projections
\[
\bP_r  =  \diag(0,0,1,0,1,0,0,0,1,0,1,1,1) ~~~\mbox{and}~~~ 
\bP_w  =  \diag(0,0,0,0,0,1,1,0,0,0,0,1,1) 
\]
then
\[
R(\lambda_{ij}) = \| \sum_{i=1}^3 (\lambda_{ij})_j \bP_r \|_1
~~\mbox{and}~~
W(\lambda_{ij}) = \| \sum_{i=1}^3 (\lambda_{ij})_j \bP_w \|_1
\]

The optimisation problem we thus have to solve is given by
\[
\min \Phi_{\rho\omega}(\la_{ij})
~~\mbox{subject to:}~~
\sum_{j} \lambda_{ij}=1 ~\forall i=1,2,3
~~\mbox{and}~
0 \leq \lambda_{ij} \leq 1 ~\forall i=1,2,3, j=1,\ldots,13
\]

Using the builtin non-linear optimisation in {\tt octave} we get for certain
initial $\la_{ij}$'s the some interesting results when we minimise the
objective function $\Phi_{\rho\omega}$.

\nocite{Octave}

If we start with a swap which uses $z$, like
\(
\getsL{z}{x}{6};~  \getsL{x}{y}{2};~ \getsL{y}{z}{5}
\) 
which corresponds to $39$ values for $\la_{ij}$ below (each row corresponds
to the three computational steps, and each column to the weight of each of the
$13$ possible blocks).%
For $\min \Phi_{11}$ we get after $12$ iterations of the standard non-linear 
optimisation algorithm in {\tt octave} a program transformation namely the 
following set of $\la_{ij}$:
\[
\left(
\begin{array}{ccccccccccccc}
0 & 0 & 0 & 0 & 0 & 1 & 0 & 0 & 0 & 0 & 0 & 0 & 0 \\ 
0 & 1 & 0 & 0 & 0 & 0 & 0 & 0 & 0 & 0 & 0 & 0 & 0 \\ 
0 & 0 & 0 & 0 & 1 & 0 & 0 & 0 & 0 & 0 & 0 & 0 & 0 \\ 
\end{array}
\right)
\mapsto
\left(
\begin{array}{ccccccccccccc}
0 & 0 & 0 & 0 & 0 & 0 & 0 & 0 & 0 & 1 & 0 & 0 & 0 \\ 
0 & 0 & 0 & 0 & 0 & 0 & 0 & 1 & 0 & 0 & 0 & 0 & 0 \\ 
0 & 0 & 0 & 0 & 0 & 0 & 0 & 0 & 0 & 1 & 0 & 0 & 0 \\ 
\end{array}
\right)
\]
This corresponds to the program:
\(
\getsL{y}{(y+x) \bmod 2}{10};~ \getsL{x}{(x+y) \bmod 2}{8};~ \getsL{y}{(y+x)
  \bmod 2}{10}
\) 
which indeed also swaps $x$ and $y$ but does not use the variable $z$ in any
way.

For randomly chosen initial values (which do not even fulfill the
normalisation conditions for $\la_{ij}$) we get with the objective function
$\Phi_{11}$ from
\[
\left( 
  \begin{array}{ccccccccccccc} 
    .70 & .30 & .72 & .84 & .51 & .70 & .76 & .47 & .63 & .63 & .93 & .55 & .68 \\
    .74 & .22 & .37 & .70 & .67 & .13 & .93 & .69 & .30 & .88 & .03 & .52 & .80 \\ 
    .59 & .49 & .01 & .69 & .22 & .23 & .10 & .01 & .10 & .22 & .03 & .55 & .11 \\ 
  \end{array} \right)
\]
after $9$ iterations with {\tt octave}:
\[
\left(
\begin{array}{ccccccccccccc}
0 & 0 & 0 & 0 & 0 & 0 & 0 & 0 & 0 & 1 & 0 & 0 & 0 \\ 
0 & 0 & 0 & 0 & 0 & 0 & 0 & 1 & 0 & 0 & 0 & 0 & 0 \\ 
0 & 0 & 0 & 0 & 0 & 0 & 0 & 0 & 0 & 1 & 0 & 0 & 0 \\ 
\end{array}
\right)
\]
This corresponds to the program:
\[
\getsL{y}{(y+x) \bmod 2}{10};~  
\getsL{x}{(x+y) \bmod 2}{8};~ 
\getsL{y}{(y+x) \bmod 2}{10}
\]
For $\Phi_{00}$ we may also get $\getsL{z}{x}{6};~ \getsL{x}{y}{2};~
\getsL{y}{z}{5}$.
Sometimes however the optimisation does not work, we get stuck in a local
minimum. This can usually be overcome by using stronger punishing terms for
read and writes, e.g. $\Phi_{100,100}$.
It might be interesting to observe that we obtain the desired program without
reference to any {\em algebraic} properties of the xor operation.


\section{Conclusions and Further Work}
\label{Conclusions}

The idea of our approach is to consider a continuous manifold of program
sketches which are parametrised using some (real-valued, probability)
parameters. The objectives of the synthesis are encoded as a global
optimisation problem in terms of the LOS semantics of these programs. The PAI
framework can often be used in order to extract the relevant properties or
information.
 
This setting allows us to address quantitative and qualitative problems within
the same setting. Initial numerical experiments also indicate that although we
allow probabilistic choices in the intermediate programs the ultimate solution
are often purely deterministic programs.

The approach is in many ways orthogonal to the one in a recent paper by
Chaudhuri et.al. \cite{ClochardEtAl14}. We also consider a probabilistic
semantics but while \cite{ClochardEtAl14}\ is in essence based on Kozen's
denotational semantics our approach utilises LOS which is more in the spirit
of a compositional small-step collecting semantics. In the presentation here
we restrict ourselves to a finite dimensional version, though it is possible,
see \cite{APLAS13}, to drop this restriction: however the result is a Hilbert
rather than a Banach space semantics (based on measure theoretic structures as
in the case of Kozen's semantics). Our semantics is compositional because it
exploits tensor products in order to describe individual effects (similar to
SAN models). We also make use of Probabilistic Abstract Interpretation as
introduced in \cite{PPDP00}\ rather than Abstract Interpretation of a
probabilistic semantics (as in the case of Monniaux's approach
\cite{Monniaux00}, and more recently in \cite{CousotMonerau12}). We also use
PAI in a different way: in order to define or extract quantitative properties
of programs rather than to construct (smooth) semantical approximations as in
\cite{ClochardEtAl14}. Our setting provides for continuous optimisation
problems because of the parametrisation of the program semantics not its
abstraction.

For the time being it is unclear if and in which way the presented approach
would scale. Problems in this respect could be overcome by the compositional
nature of our semantics, but additional techniques will be needed to
accelerate the optimisation process which could include for example: guided
search using ideas from differential (information) geometry or exploiting the
structure, symmetry etc. of a particular problems via PAI in maybe a similar
way as classical AI has been used in the synthesis of synchronisation in
\cite{VechevEtAl10}; it could also be useful to combine this with a staged
optimisation, addressing more detailed constraints after more global ones have
been resolved, e.g. incorporating elements of \cite{ClochardEtAl14}. It would
also be desirable to combine our framework with formal specifications (logics)
of program properties and synthesis objectives. Finally, the experimental
tools used up to now should be developed into more efficient and user-friendly
versions.



\bibliographystyle{eptcs}
\bibliography{SYNT14}


\end{document}
